\pdfoutput=1   

\documentclass[9pt,twocolumn,twoside]{osajnl}

\journal{ao} 

\setboolean{shortarticle}{false}

\usepackage{amssymb}
\usepackage{amsmath}
\usepackage[separate-uncertainty=true,multi-part-units=single,range-units=single]{siunitx}
\usepackage{subfig}

\renewcommand{\vec}[1]{\mathbf{{#1}}}                         
\newcommand{\pvec}[1]{\mathbf{{#1}}_\parallel }
\newcommand{\vecUnit}[1]{\mathbf{\hat{#1}}}                         
\newcommand{\pvecUnit}[1]{\mathbf{\hat{#1}}_\parallel }

\newcommand{\enquote}[1]{``{#1}''}

\usepackage{xcolor}

\renewcommand{\eqref}[1]{(\ref{#1})}

\graphicspath{.,{Figures/}}         

\title{Experimental studies of the transmission of light through low coverage regular or random arrays of silica micropillars supported by a glass substrate}

\author[1]{Colette Turbil}
\author[2]{Thomas Sang HyukYoo}
\author[1,3]{Ingve Simonsen}
\author[1]{Jeremie Teisseire}
\author[1,*]{Iryna Gozhyk}
\author[2]{Enric Garcia-Caurel}

\affil[1]{SVI, UMR 125 CNRS/Saint-Gobain Recherche, 39 quai Lucien Lefranc, 93303 Aubervilliers, France}
\affil[2]{LPICM, CNRS, Ecole Polytechnique, Université Paris-Saclay, 91128 Palaiseau, France}
\affil[3]{Department of Physics, NTNU -- Norwegian University of Science and Technology, NO-7491 
Trondheim, Norway}

\affil[*]{Corresponding author: iryna.gozhyk@saint-gobain.com}

 \begin{abstract}
   The transmission of light through low coverage regular and random arrays of glass supported silica micropillars of diameters \SIrange{10}{40}{\micro\meter} and height \SI{10}{\micro\meter} is studied experimentally. Angle-resolved measurements of the transmitted intensity are performed at visible wavelengths by either a goniospectrophotometer or a multimodal imaging (Mueller) polarimetric microscope. It is demonstrated that for the regular arrays, the angle-resolved  measurements are capable of resolving many of the densely packed diffraction orders that are expected for periodic structures of lattice constants \SIrange{20}{80}{\micro\meter}, but they also display features that are due to the scattering and guiding of light in individual micropillars or in the supporting glass slides. These latter features are also found in angle-resolved measurements on random arrays of micropillars of the same surface coverage. Finally we perform a comparison of direct measurements of haze in transmission for our patterned glass samples with what can be calculated from the angle-resolved transmitted intensity measurements. Good agreement between the two types of results are found which testifies to the accuracy of the angle-resolved measurements that we report.   
 \end{abstract}

\begin{document}

\dates{} 
\ociscodes{(050.1940) Diffraction; (120.0120) Instrumentation, measurement, and metrology; (240.0240) Optics at surfaces; (290.0290) Scattering}
\doi{}
\maketitle


\section{Introduction}
\label{Sec:Intro}
Surfaces with features from macro~\cite{Candela} to nanoscale~\cite{Bhushan} can be found both in
nature and in industrial products. Such features cause light scattering and diffraction
and, therefore, they impact the visual appearance of objects. Among the phenomena related to light
scattering and diffraction there are many non-exotic effects which can be experienced in
everyday life: reflection of light from, or transmission trough, a rough surface~\cite{Stover,Simonsen2010};
diffraction through apertures~\cite{BornWolf}~(holes in the window blinds) or from or through
a transparent/opaque material of a particular shape (Mie scattering~\cite{Cohen1982}); diffraction
from periodic structures (diffraction gratings~\cite{Loewen1997}, photonic crystals~\cite{Joannopoulos}).
In textbooks these phenomena are often considered a separate phenomenon  and often restricted to a
particular range of parameters. For instance, ``diffraction grating'' analysis is almost
exclusively applied to periodic structures for which the lattice constants are comparable to the
wavelength of the incident light, which for visible light, corresponds to roughly \num{0.5}--\SI{2}{\micro\meter}.
Meanwhile, the habit of considering only specific cases of light scattering and diffraction for a particular range of
parameters may lead to an incorrect approach to the characterization of samples and thus to inaccurate or even incorrect 
interpretation of experimental data.

Periodic patterns with lattice constants from tens to hundreds of micrometers exhibit hundreds
to thousands of (propagating) diffraction orders when the sample is illuminated by visible light. Such structures are by no
means efficient diffraction gratings. Yet, as will be shown in this paper, they can still produce a
noticeable effect on the optical properties of the samples and thus their visual appearances. Moreover,
if several feeble optical effects are brought together they can impact significantly the optical
response of the surfaces.

Measurements of gloss and haze are often the first step in the experimental characterization of the
optical properties of samples which scatter light due to surface patterns or volume disorder. The simple
integral optical properties --- haze and gloss --- allow a distinction of the light that is reflected/transmitted specularly
by the sample  from the light that is scattered diffusely by it.
However, the actual spectral and angular distributions of the intensity of the scattered light can be rather
complex. Despite the convenience of haze and gloss in terms of the simplicity of measurement and speed,
such integral properties may misguide the interpretation, especially if such measurements 
contain the artifacts related to the diffraction of light.

\smallskip
The rest of this paper is organized in the following way. Section~\ref{Sec:Methods} presents the samples that we will study in this work and the instruments used to perform the angle-resolved intensity measurements. Such and other results are presented in Sec.~\ref{Sec:Results} where we also discuss and interpret the various features that are present in them. In particular, in this section, we discuss origin of the different types of circular intensity fringes that the angle-resolved transmitted intensity distributions possess. The haze of the samples, both directly measured, and obtained on the basis of the angle-resolved measurements are also discussed here [Sec.~\ref{Sec:Results}.\ref{Sec:Haze}]. Finally the conclusions that can be drawn from this work are presented in Sec.~\ref{Sec:Concusions}. 
  
\section{Materials and Methods}
\label{Sec:Methods}

\subsection{Description of the samples}
\label{Sec:samples}
\begin{figure}[tbhp]
  \begin{center}
    \centering
    \includegraphics[width=0.6\columnwidth]{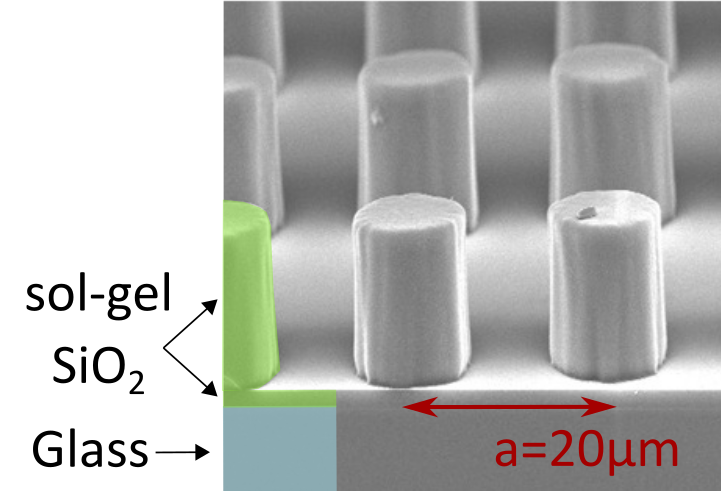}
    \caption{Scanning electron microscope image of a cross-section of the sample taken at grazing incidence. The height of all micropillars used in this study was $h=\SI{10}{\micro\meter}$. }
    \label{Fig:2}
  \end{center}
\end{figure}

The studied samples consist of \SI[product-units=power]{5 x 5}{\cm} glass slides covered by an array of cylindrical silica micropillars [Fig.~\ref{Fig:2}]; all cylinders had the height $h=\SI{10}{\micro\meter}$ while their diameter were $d=\SI{10}{\micro\meter}$, \SI{20}{\micro\meter} or \SI{40}{\micro\meter}. Both regular and random arrays were studied. The regular arrays were either hexagonal or square, for which the lattice vectors are $\vec{a}_1= a\vecUnit{x}$ [both cases] and $\vec{a}_2 = (a/2)[-\vecUnit{x} + \sqrt{3}\vecUnit{y}]$~[hexagonal] or $\vec{a}_2 = a \vecUnit{y}$~[square], where $\vecUnit{x}$ and $\vecUnit{y}$ are orthogonal unit vectors in the sample plane (the $xy$-plane). The lattice constant $a$ of the regular arrays varied from \SIrange{20}{80}{\micro\meter}~[Fig.~\ref{Fig:2}]. The surface coverage $\rho$ is defined as the ratio between the areas of the base of one micropillar to the area of a unit cell. If the diameter of the micropillars is denoted $d$, one finds that for a hexagonal array  
$\rho = (\pi/2\sqrt{3})(d/a)^2$ while for a square array $\rho=(\pi/4)(d/a)^2$~\cite{Turbil2016}.
For instance, for pillars of diameter  \SI{10}{\micro\meter} arranged in a hexagonal array the previously given values of $a$ correspond to the surface coverage between $\rho=\SI{1.4}{\percent}$ ($a=\SI{80}{\micro\meter}$) and \SI{22.7}{\percent} ($a=\SI{20}{\micro\meter}$). Random arrays were produced to have the same surface coverage as the regular arrays for the same type of micropillars. To this end, the centers of the pillars were chosen to be uniformly distributed in such a way that two pillars could not be closer to each other than a minimum center-to-center distance (\SI{20}{\micro\meter}). About a \SI{300}{nm} thick residual silica layer between the glass surface and the base of the pillars provided an adhesion layer for the pillars to the surface of the substrate~[Fig.~\ref{Fig:2}]. The thickness of the glass substrates was \SI{2}{mm} for the majority of samples, and \SI{1}{mm} for samples measured with the multimodal microscope. All the samples were produced in sol-gel silica by the use of  the nanoimprinting technique (for detailed description see Ref.~\cite{Dubov2013}).

\subsection{Optical measurements}

Conventional integral haze and gloss measurements were used in this study together with two
facilities for angle-resolved measurements: OMS4 goniospectrophotometer (commercialized
by OPTIS) and a home-made multimodal imaging (Mueller) polarimetric microscope. 
Both angle-resolved setups allow for the measurements of the optical intensity response of the
sample. Since the two instruments are based on very different optical configurations, we
decided to use them both to verify the consistency of the results and to identify the presence
of possible artifacts due to the instrument response functions. Despite of the fact that the two
instruments also allowed for the measurement of the state of polarization of the transmitted or
reflected light~\cite{Yoo2017}, this paper focuses on the angular distribution in the far field of the intensity
transmitted through the sample regardless of its polarization.

The integral optical properties, haze and gloss, were obtained by the use of a 
\textit{hazemeter} (BYK Gardner Haze-gard plus) for measurements in transmission and a
\textit{glossmeter} (Enrichsen -- Pico Glossemaster Model 500) for measurements in reflection.

Being a goniospectrophotometer device, OMS4 consists of the sample holder and two
arms where one of the arms have a set of light sources installed and the other a photomultiplicator detector. The
sample holder and detector arm can be moved automatically with the help of one and two
precise motors, respectively. This allows to scan the whole angular region around the sample and thus,
to measure the angular intensity distributions of the reflected or transmitted light, at any angle of incidence in the
range of $\theta_i\in[\ang{0}; \ang{85}]$.

In this study OMS4 was used in transmission mode to  measure the angular distribution of the transmitted intensity; the angular resolution was \ang{0.5} around the specular direction. Measurements were performed with three coherent laser sources~(RGB) and an incoherent Xenon lamp light with or without color filters. The bidirectional
transmittance distribution function~(BTDF)~\cite{Bartell1981} of the samples was collected for
the polar angles of incidence $\theta_i=\ang{0}$, \ang{10}, \ang{30} and \ang{60}. A related function, the differential transmission coefficient~(DTC)~\cite{Hetland}, is obtained by multiplying the BTDF by the cosine of the polar angle of transmission~[$\cos\theta_t$].
The contour plots of the DTC, obtained in this way from data measured by OMS4, were plotted with the SPEOS software package produced by OPTIS.

\begin{figure}[tbhp]
  \begin{center}
    \centering
    \includegraphics[width=0.85\columnwidth]{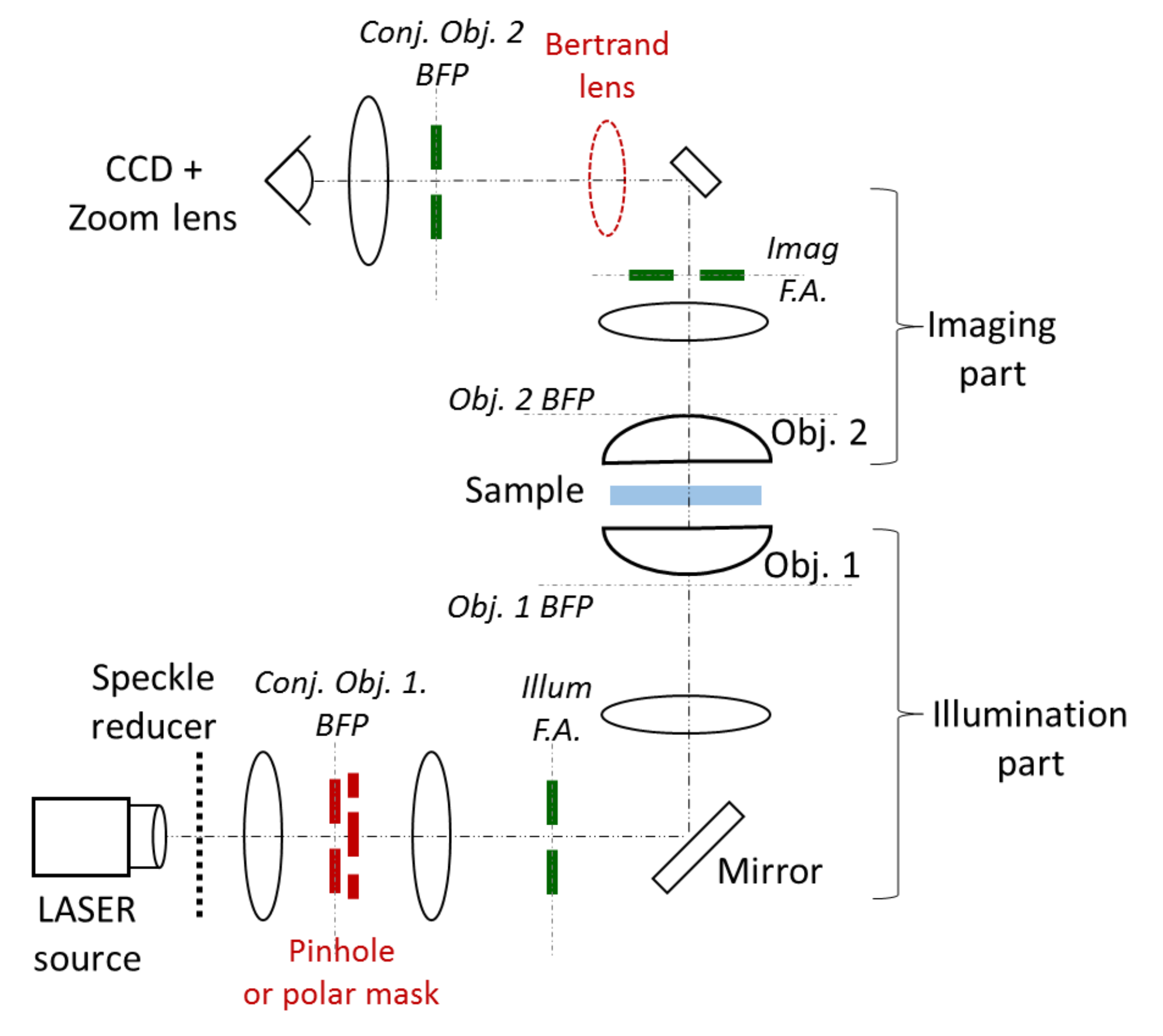}
    \caption{Schematic illustration of the multimodal (Mueller) microscope in the
      transmission configuration. The positions of the conjugate images corresponding to the back-focal
      planes~(BFP) of the two objectives, as well as the conjugate planes of the sample in the
      illumination and imaging arm are shown. Also indicated are the positions of the retractable Bertrand lens, the
      light source, and the detection camera.}
    \label{Fig:3}
  \end{center}
\end{figure}

\smallskip
The main difference between the goniospectrophotometer and the multimodal
polarimetric microscope is the absence of moving parts in the latter system. The multimodal
microscope can be operated in two imaging modes, real plane and Fourier (or conjugate
space) plane imaging modes. In real plane imaging mode the microscope produces images of the studied
sample, while in Fourier imaging mode the images correspond to the angular distribution of
light transmitted or reflected by the sample. The optical configuration of the multimodal
microscope is sketched in Fig.~\ref{Fig:3}. The instrument was coupled to a laser emitting green light
at a wavelength of \SI{533}{nm} with a spectral width of less than \SI{2}{nm}. Speckle effects due to the
coherence of the laser were minimized using a vibrating rough membrane (Laser Speckle
reducer from Optotune) just in front of the laser source. The microscope was mounted in
the transmission configuration; the sample was located between two identical microscope
objectives (one for imaging and another one for illumination). The microscope objectives can be
selected to have different magnifications; $50\times$, $20\times$, $10\times$, and $5\times$
depending on the required resolution and a numerical aperture.

Thanks to the use of a series of relay lenses, it is possible to create conjugate images of the
back-focal planes (BFP) of the objectives in both the illumination and imaging parts.
Therefore, we can insert apertures in the conjugate plane of the BFP of the illuminating
objective with different shapes and sizes to simultaneously control the direction and the
angular aperture of the illuminating beam. Analogously, the insertion of a pinhole or polar mask
at the conjugate plane of the BFP of the imaging microscope objective, allows controlling the
direction and aperture of the detected scattered beam.

The direction of the illuminating beam is defined by the mean polar angle of incidence
$\left<\theta_i\right> = \arcsin(D/f)$.
Here $f$ is the focal length of the microscope objective and  $D$ the off-axis distance measured from the
center of the pinhole to the optical axis of the microscope. For instance, if a pinhole is placed in the plane conjugated to the illuminating objective BFP,
and, this same pinhole is shifted to a given distance to the optical axis, then the sample can be
illuminated with an oblique incidence. When the pinhole is aligned with the microscope
optical axis, the sample is illuminated at normal incidence.

Moreover, once the average polar angle $\left<\theta_i\right>$  is known, the divergence (div in radians) of the
illumination, or alternatively the imaging beam, can be expressed as a function of the
corresponding pinhole diameter $\phi_{\textrm{pin}}$, the focal length of microscope objective $f$, and the mean
polar angle according to:
\begin{align}
 \textrm{div}
  &=
    \frac{\phi_{\textrm{pin}}}{f \cos \left<\theta_i\right>}.
  \label{eq:6}
\end{align}
The relay lens system also provides a conjugate of the object plane (the sample) in both; the
illumination and the imaging arms, therefore, the use of pinholes or polar masks in those
planes, helps to define the shape and size of the illuminated and imaged area of the sample,
or, in other words, the field of view (FOV). The insertion of a Bertrand lens in the optical
path of the microscope allows to easily switching between the real and the Fourier imaging
modes~\cite{Kurvits2015}.

\section{Results and discussions}
\label{Sec:Results}

First in this section we are concerned with angle-resolved measurements for both the regular and the random arrays of micropillars and the discussion of the features that such data show. Later we address the integral optical properties of the samples, like haze and gloss. This will be done by direct measurements of haze, but also by calculations based on the angle-resolved measurements that we performed.

\subsection{Angle-resolved measurements}
\label{SubSec:3B}

In standard scatterometry measurements, such as BTDF or DTC measurements, the direction of illumination
and detection are both selected by moving the goniometric arms on which the source and the
detector are mounted, respectively. In the multimodal microscope, however, the control of the
direction of the beam in both the illumination and the imaging arm is obtained by the use of pinholes
placed at different positions in the conjugate planes~(BFP) of the microscope objective as
previously discussed. These intrinsic differences in how the measurements are performed
with these two setups make a comparison of the obtained data very interesting.

\begin{figure}[tbhp]
  \begin{center}
    \centering
    \includegraphics[width=0.8\columnwidth]{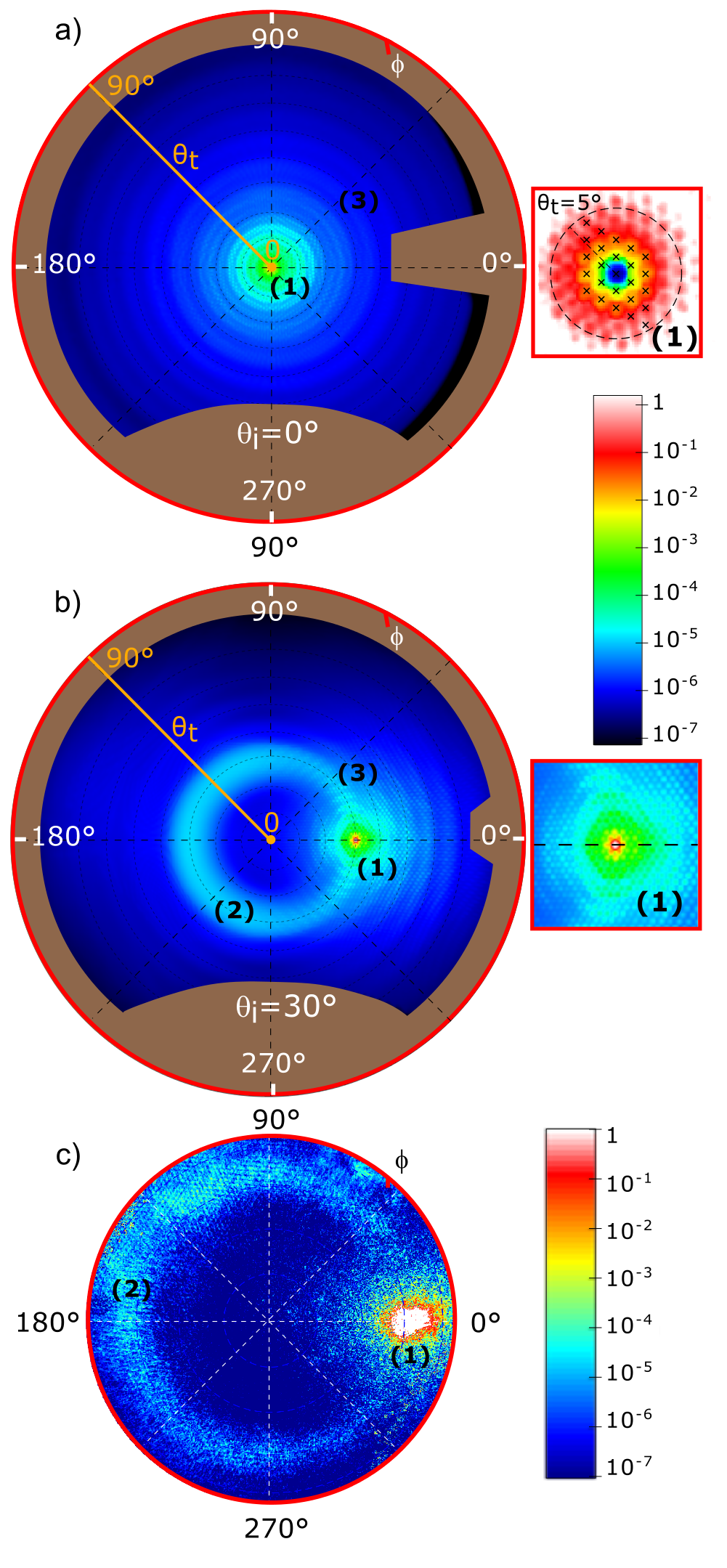}
    \caption{
      The angular distribution of the DTCs,  normalized by their maximum values, and measured with
      (a)~a goniospectrophotometer and (b)~a multimodal microscope in the Fourier plane imaging mode.
      In both cases, the polar angle of incidence was $\theta_i=\ang{30}$ and the sample consisted of a hexagonal array of cylindrical micropillars
      of diameter $d=\SI{10}{\micro\meter}$, height $h=\SI{10}{\micro\meter}$, and the lattice constant was $a= \SI{30}{\micro\meter}$.
      The measurements were performed at wavelengths (a)~\SI{535}{nm} and (b)~\SI{533}{nm}. In panel~(b) the angular aperture of the
      illuminating beam was \ang{10}. The angular distribution of the measured DTCs contain three main features (as marked in each of the panels):
      (1)~a specular peak and diffraction orders; (2)~a diffuse ring; and (3)~concentric patterns.
      The inset in panel~(a) shows the details around the specular direction of an angular width of $\pm\ang{20}$. The regions of polar angles of transmission over which the measurements were performed were $\theta_t\leq\ang{85}$~[Figs.~\ref{Fig:5}(a)--(b)] and $\theta_t\leq\ang{45}$~[Fig.~\ref{Fig:5}(c)].}
    \label{Fig:5}
  \end{center}
\end{figure}

Figure~\ref{Fig:5}  presents contour plots of the angular distribution of the normalized DTCs obtained for a sample of a hexagonal array of micropillars and measured by either the goniospectrophotometer[Figs.~\ref{Fig:5}(a)--(b)] or the microscope [Fig.~\ref{Fig:5}(c)]. The normalization was done with respect to the maximum value of the angular dependent DTC, which in our case, was found in the direction of specular transmission.

We first start by discussing the measurements performed with the  goniospectrophotometer. In the outer brown region of Figs.~\ref{Fig:5}(a) and ~\ref{Fig:5}(b) no measurements were performed since either the detector arm covered for the source (around $\phi_t=\ang{0}$ in  Fig.~\ref{Fig:5}(a)) or the angular region being inaccessible to the detector due to the physical dimensions of the support on which the setup is mounted (around $\phi_t=\ang{270}$ in Figs.~\ref{Fig:5}(a) and \ref{Fig:5}(b)). The specular direction of transmission in Fig.~\ref{Fig:5}, and in the preceding experimental results to be presented, is at $(\theta_t, \phi_t)=(\theta_i,\phi_i-\ang{180})=(\theta_i,\ang{0})$. The sample consisted of cylindrical pillars of diameter $d=\SI{10}{\micro\meter}$ and the lattice constant was $a=\SI{30}{\micro\meter}$~[$\rho=\SI{10}{\percent}$]. The angles of incidence assumed in obtaining the results in Fig.~\ref{Fig:5}(a) were $(\theta_i,\phi_i)=(\ang{0},\ang{180})$ while in Figs.~\ref{Fig:5}(b)--(c) the angles of incidence were $(\theta_i,\phi_i)=(\ang{30},\ang{180})$. In obtaining the results in
Figs.~\ref{Fig:5}(a)--(b) the illuminating source consisted of a Xenon lamp to which  a \SI{10}{nm}-wide spectral filter centered at \SI{535}{nm} was applied; a laser source of wavelength $\lambda=\SI{533}{nm}$ was used to obtain the results presented in Fig.~\ref{Fig:5}(c). All the transmitted light, independent of polarization, was detected. It is challenging to accurately align the micro-patterned sample in the macroscopic setup so therefore the azimuthal angle of incidence may show slight deviations from $\phi_i=\ang{180}$. It should be mentioned that for the  case of regular arrays of micropillars we for all experiments tried to align the samples so that the plane of incidence contains the lattice vector $\vec{a}_1$.

Several interesting features should be observed from the results presented in Fig.~\ref{Fig:5}. First, specular
peaks of high transmitted intensity are observed in the measurements at $(\theta_t,\phi_t)=(\theta_i, \ang{0})$,
and the transmitted intensities drop off away from this direction.

Second, the measurements have sufficient angular resolution to allow for the observation of the dense pattern of propagating diffractive orders; this is most apparent from Fig.~\ref{Fig:5}(a) and the region marked~(1) in Fig.~\ref{Fig:5}(b) where the diffraction orders appears like small dots.
The positions of the propagating diffractive orders are determined by the grating equation which states that the lateral wavevector of transmission is given by
\begin{align}
  \pvec{q}^{(m)} = \pvec{k}+\vec{G}_m,
  \label{eq:grating_equation}  
\end{align}
where $\vec{G}_m=m_1\vec{b}_1+m_2\vec{b}_2$ is a vector defined in terms of the primitive lattice vectors $\vec{b}_i$  of the reciprocal lattice, and $m=\{m_1,m_2\}$ denotes a set of two integers~[$m_i\in\mathbb{Z}$] defining a diffractive order. The primitive lattice vectors $\vec{b}_j$ are obtained from the relations $\vec{a}_i\cdot \vec{b}_j= 2\pi \delta_{ij}$. For a hexagonal array of lattice vectors $\vec{a}_i$ defined in Sec.~\ref{Sec:Methods}.\ref{Sec:samples}, we obtain the primitive lattice vectors  $\vec{b}_1=(2\pi/a)[\vecUnit{x}+\vecUnit{y}/\sqrt{3}]$ and $\vec{b}_2=(2\pi/a)[2 \vecUnit{y}/\sqrt{3}]$.
The  wavevector $\pvec{q}^{(m)}$ of the diffractive order characterized by $m$ and the angles $(\theta_t^{(m)},\phi_t^{(m)})$ is defined in terms of the angles of transmission as $\pvec{q}=(\omega/c)\sin\theta_t(\cos\phi_t,\sin\phi_t,0)$ with $(\theta_t,\phi_t)=(\theta_t^{(m)},\phi_t^{(m)})$ where  $\omega/c=2\pi/\lambda$ is the vacuum wave number. The wavevector $\pvec{k}$ of the incident light is defined in a similar manner but in terms of the angles of incidence $(\theta_i,\phi_i-\ang{180})$. 
The polar angles of transmission $\theta_t^{(m)}$ for the diffractive order characterized by $m$ is therefore obtained from the relation $q_\parallel^{(m)}=(\omega/c)\sin\theta_t^{(m)}$ and the corresponding azimuthal angle of transmission from the orientation of the unit vector $\pvecUnit{q}^{(m)}$. 
In this way we calculated the angular positions of the various diffractive orders indicated in Fig.~\ref{Fig:5}(a) as black dots; this figure displays a good agreement between the measured and the theoretically predicted angular positions of the diffractive orders.

Third, and probably most unexpected and interesting, are the additional complex angular patterns that are visible in the DTC for non-normal incidence~[Figs.~\ref{Fig:5}(b)--(c)] and which are further away from the specular directions. These features are marked~(2) and (3) in Fig.~\ref{Fig:5}(b). The first of these features, refers to  the \textit{circular structure} of high transmitted intensity seen in this figure. In the following we will simply refer to it as a ``halo''. The circular halo is centered at the direction of normal transmission~[$\theta_t=\ang{0}$]  and it contains the direction of specular transmission $(\theta_t,\phi_t)=(\theta_i,\ang{0})$. This means that the angular position of the halo is defined by the equation $\theta_t=\theta_i$ (which is also the  angular radius of the circular structure). This has the consequence that the halo is predicted to disappear, or coincide with the specular, for light that is incident normally onto the micropillar array. 
That this is indeed the case, can be observed from Fig.~\ref{Fig:5}(a), where the measured data do not show any apparent signs of a halo. Further discussion and details about the halo will be provided in Sec.~\ref{Sec:Results}.\ref{SubSec:3B1}.

The features marked~(3) in Fig.~\ref{Fig:5}(b) refer to the ``intensity oscillations'', or \textit{fringes}, observed in the angular distribution of the DTC  in the angular region around the specular direction that also is outside the halo. As will be discussed in greater detail in Sec.~\ref{Sec:Results}.\ref{SubSec:3B3}, these fringes belong to two distinct classes of \textit{concentric patterns}: (i)~those circular fringes that are concentric to the specular direction of transmission and (ii) those  patterns that are concentric to the normal direction of transmission $(\theta_t,\phi_t)=(\ang{0},\ang{0})$ [the ``origin'']. The former class of fringes are, for instance, readily observed within a \ang{10} cone around the specular direction of transmission in Figs.~\ref{Fig:5}(a)--(b), while the latter class is particularly visible for larger polar angles of transmission Fig.~\ref{Fig:5}(a).

\smallskip
We now turn to the measurement performed by the multimodal microscope. Figure~\ref{Fig:5}(c) presents the result for the angular distribution of the normalized DTC obtained in this way when light was incident at the polar angle of incidence $\theta_i = \ang{30}$ on a hexagonal array of micropillars of the same kind as used to produce the results in Figs.~\ref{Fig:5}(a)--(b).
The microscope measurement was performed for the range of polar angles of transmission $\theta_t\leq \ang{45}$ and all azimuthal angles of transmission $\phi_t$ with no aperture restriction imposed in the imaging arm.  Moreover, the wavelength of the incident light was $\lambda = \SI{533}{nm}$. This is \SI{2}{nm} less than the wavelength used to perform the measurements with the goniospectrophotometer, but we believe that this minor difference in wavelength will not cause any significant difference in the obtained results. The four concentric dashed circles seen in Fig.~\ref{Fig:5}(c) indicate the polar angles of transmission $\theta_t=\ang{10}$, \ang{20}, \ang{30} and \ang{40} (from inner to outer). The white circular area in this figure represents the region around the specular direction.

The angular aperture, $\Theta$, or in other words, the maximum polar angle that can be measured by the multimodal microscope is $\sim\ang{45}$. This value is given by the numerical aperture~(NA) of the objective of the microscope used, which in our case was identical to the  magnification $50\times$ and an NA of \num{0.85}. In this case the minimum available thickness of the glass substrate on which the cylinders were deposited was \SI{1}{mm}. In consequence, using objectives with higher NA and thereby working distance much smaller than \SI{1}{mm}, was not possible in this study.

The angular dependence of the normalized DTC measured with the multimodal microscope is, for the same angles of incidence, 
similar to the same quantity  measured with the goniospectrophotometer.
An additional analogy between the images taken with the goniospectrophotometer
and the multimodal microscope is the fact that the diffracted orders are also visible in the
image taken with the multimodal microscope as an ensemble of regularly spaced spots (light
blue) clearly seen on a dark blue background.
In the image taken with the goniospectrophotometer the intensity oscillations that are due to scattering are easily seen to be located for polar angles of transmission  $\theta_t >\theta_i$ (on the right of the specular spot). Unfortunately, in the image taken by the multimodal microscope, this angular region of high contrast falls outside the measured area because the polar angle of the incident beam is close to the maximum aperture accepted by the microscope objective used for the measurements.

\subsubsection{Diffraction orders due to the regular array of pillars}
\label{SubSec:3B1}

\begin{figure}[tbhp]
  \begin{center}
    \centering
		\includegraphics[width=0.99\columnwidth]{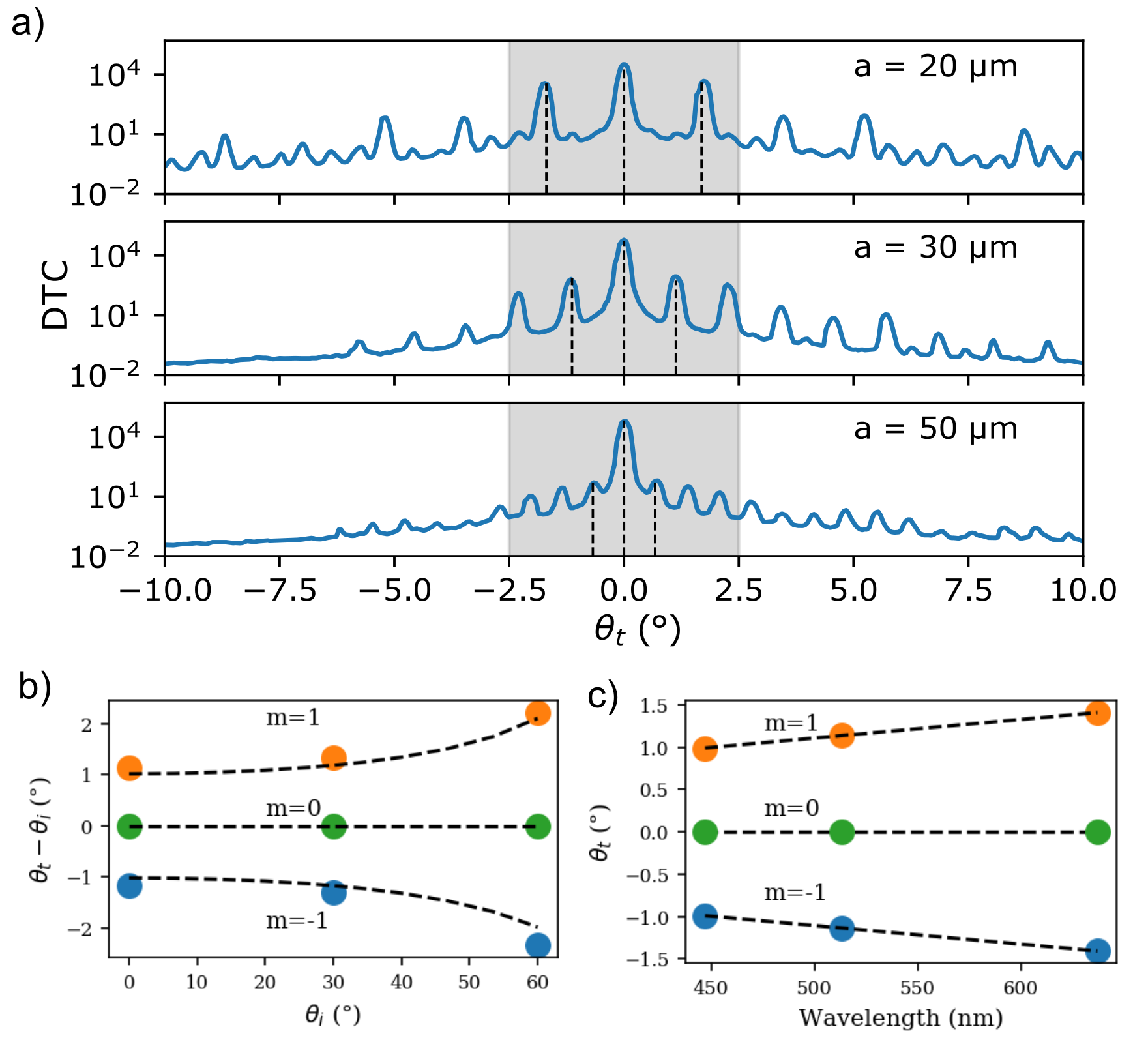}
        \caption{The out-of-plane dependence of the DTCs measured by the goniospectrophotometer for hexagonal arrays of micropillars ($d=\SI{10}{\micro\meter}$) illuminated by a laser source of wavelength $\lambda$. 
          (a) The angular dependence of the out-of-plane DTCs for a set of three values of the lattice constants $a$ for $\theta_i=\ang{0}$ and $\lambda=\SI{513}{nm}$;
          the dependence of the polar angle of transmission $\theta_t^{(m)}$ for $m=\{0,m_2\}$ (out-of-plane) obtained experimentally (filled symbols) as a function of (b)~the polar angle of incidence $\theta_i$ for $\lambda=\SI{513}{nm}$; or   (c)~the wavelength $\lambda$ for normal incidence [$\theta_i =\ang{0}$]. In panels~(b) and (c) the lattice constant was $a=\SI{30}{\micro\meter}$ and $m_2=0,\pm 1$.
          The plane of incident was chosen so that the lattice vectors $\vec{a}_1$ lies in this plane.
      The dashed black lines represent the theoretic values for the polar angle of diffraction $\theta_t^{(m)}$ obtained on the basis of Eq.~\protect\eqref{eq:grating_equation}. The shaded gray areas in panel~(a) represent the regions for which $\theta_t\leq \ang{2.5}$. } 
    \label{Fig:6}
  \end{center}
\end{figure}

Now we will more closely inspect the intensities and positions of some of the propagating diffractive orders that the hexagonal arrays of micropillars give rise to. Such results will in part provide information on the quality and consistency of the measurements. The results reported in Fig.~\ref{Fig:5} where obtained using narrow band sources. Therefore, to facilitate a more direct comparison of the measured results to those predicted from the grating equation, Eq.~\eqref{eq:grating_equation}, and to study the dependence on the wavelength of the incident light, we performed measurements in the plane of incidence by the use of a laser source of wavelength $\lambda$ in vacuum.
Figure~\ref{Fig:6}(a) presents the dependence of the out-of-plane transmitted intensity measured with the goniospectrophotometer for hexagonal arrays defined by three values of the lattice constant $a$ for normal incidence [$(\theta_i,\phi_i)=(\ang{0},\ang{0})$] and wavelength $\lambda=\SI{513}{nm}$. The angular positions of the diffractive peaks are in good agreement with the predictions from the grating equation~\eqref{eq:grating_equation} indicated by the black vertical dashed lines (for $m=\{0,m_2\}$ with $m_2=0,\pm 1$) in the figure. Furthermore, the gray shaded region represents the $\theta_t\leq\ang{2.5}$ used in the definition of haze.
Also good agreement is obtained between the measured data for $a=\SI{30}{\micro\meter}$ and the result obtained from standard diffraction theory when the polar angle of incidence is changed~[Fig.~\ref{Fig:6}(b)] or wavelength of the incident light is changed~[Fig.~\ref{Fig:6}(c)]. The latter results presented in these figures as black dashed lines are obtained on the basis of Eq.~\eqref{eq:grating_equation}.
Based on the results in Fig.~\ref{Fig:6} [and Fig.~\ref{Fig:5}(a)] we conclude that our angular dependent measurements are able to resolve well the densely packed diffractive orders of the regular arrays give rise to; at least this was the case for the values of the lattice constant $a\leq \SI{80}{\micro\meter}$ that we considered.

\subsubsection{The diffuse ring or halo: impact of the polar angle of incidence}

\begin{figure}[tbhp]
  \begin{center}
    \centering
    \includegraphics[width=0.75\columnwidth]{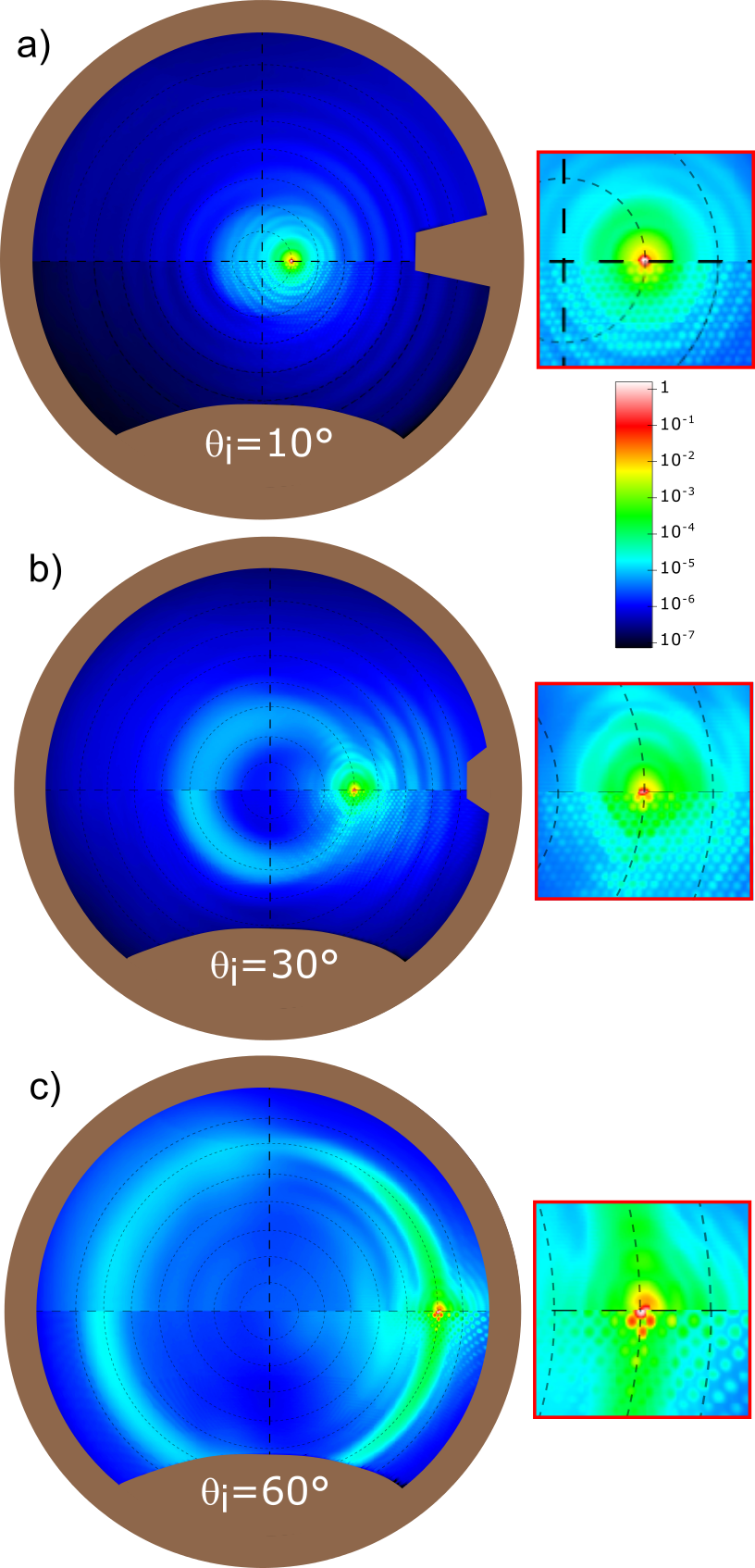}
    \caption{The angular dependence of the DTCs for the polar angles of incidence (a)~$\theta_i=\ang{10}$, (b)~\ang{30} and (c)~\ang{60} measured with the  goniospectrophotometer and normalized by their maximum values (found in the specular direction). The insets zoom in on $\pm\ang{15}$ angular regions around the specular directions. The samples consisted of a random array of micropillars~(upper halfs) or a hexagonal regular array of micropillars~(lower halfs). All micropilars were characterized by a diameter and a height of $d=\SI{10}{\micro\meter}$ and $h=\SI{10}{\micro\meter}$, respectively. The surface coverage of the two arrays were approximately \SI{10}{\percent} which for the regular array corresponds to the lattice constant of $a=\SI{30}{\micro\meter}$. In the case of the regular array, one aimed for the plane of incidence to be aligned with one of the lattice vectors $\vec{a}_i$ ($i=1,2$). The source of the incident light was a Xenon lamp filtered around the center wavelength $\lambda=\SI{535}{nm}$ by a window of width \SI{10}{nm}. The dashed grid circles are placed at multiples of $\theta_t=\ang{10}$.} 
    \label{Fig:8}
  \end{center}
\end{figure}

Next we turn to a discussion of the physical origin and the properties of the halo that is so distinctly present in Fig.~\ref{Fig:5}. To this end, we first investigate if the halo is exclusive to the regular arrays of micropillars or if it also is present in the light transmitted through random arrays of micropillars of similar surface densities. Second, we study how the intensity at the position of the halo $\theta_t=\theta_i$ varies with azimuthal angle when we change the polar angle of incidence $\theta_i$. Figure~\ref{Fig:8} presents contour plots for the angular distribution of the DTCs, measured with the goniospectrophotometer, and normalized by their maximum values for a set of polar angles of incidence. The samples were patterned by either a random array of micropillars  or a hexagonal array of micropillars. All the  micropillars were assumed to be identical and characterized by the diameter $d=\SI{10}{\micro\meter}$. Both samples were produced to have a surface coverage of about \SI{10}{\percent} which for the an hexagonal array corresponds to the lattice constant $a=\SI{30}{\micro\meter}$. The source used to illuminate these samples was light from a Xenon lamp, filtered at the center wavelength $\lambda=\SI{535}{nm}$ by a window of \SI{10}{nm} spectral width, that was unpolarized. The subplots of Fig.~\ref{Fig:8} correspond to the polar angles of incidence $\theta_i=\ang{10}$, \ang{30} and \ang{60} and in all cases  the azimuthal angle of incidence was $\phi_i=\ang{180}$~(top-down in Fig.~\ref{Fig:8}). As expected, the measured data show a reflection symmetry with respect to the plane of incidence. Therefore, for reasons of convenience, we have combined the presentation of the results for the random and regular arrays for the same surface coverage. This is done in such a way that the upper halves of each of the subplots in Fig.~\ref{Fig:8} [$\ang{0}<\phi_t<\ang{180}$] represents the normalized DTCs for the random arrays, while the lower halves [$\ang{180}<\phi_t<\ang{360}$] present the corresponding results for the regular array. The various insets in Fig.~\ref{Fig:8} present the details in an angular region around the direction of specular transmission. The first striking observation to be made from the results presented in Fig.~\ref{Fig:8} is how similar the angular intensity distributions for the random and the regular arrays of micropillars are. Of course, the transmitted intensity distributions for the random array do not display diffractive orders as is the case for the intensity distributions for the regular array. Hence, it is the intensity envelopes of the latter data sets that are similar to the corresponding intensity distributions obtained for the random array. 

\begin{figure}[tbhp]
  \begin{center}
    \centering
    \includegraphics[width=0.99\columnwidth]{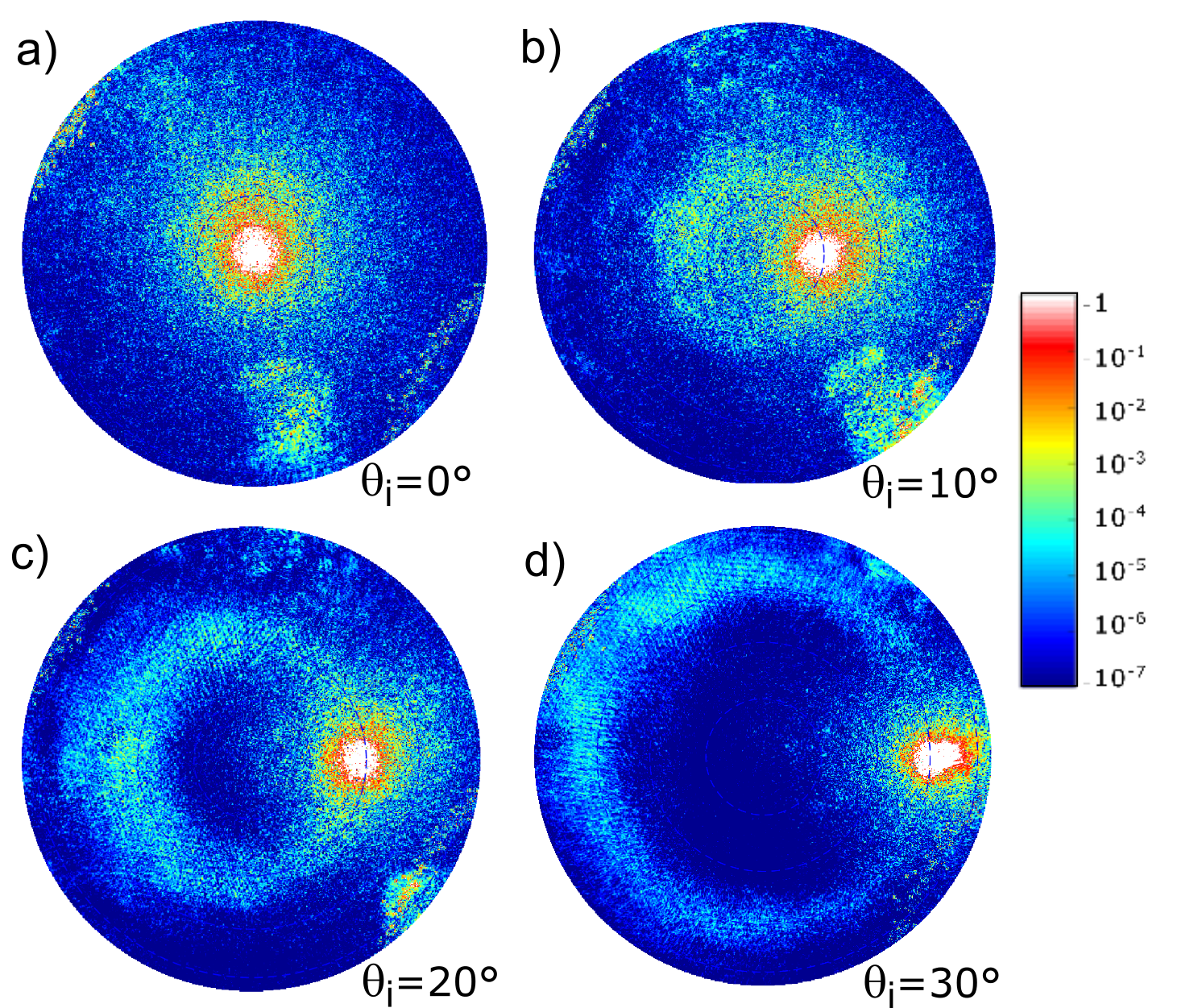}
    \caption{The far-field distribution of the normalized total intensity transmitted
      through a glass substrate patterned with a random array of micropillars of a  surface coverage of \SI{10}{\percent}. 
      The measurements were performed by the multimodal microscope in the in Fourier imaging mode.
      The polar angle of incidence of the illuminating beam was assumed to be 
      (a) $\theta_i=\ang{0}$; (b) \ang{10}; (c) \ang{20}; and (d) \ang{30}.
      The angular aperture of the illumination beam was set to \ang{10} in all cases and the measurements were performed for the wavelength $\lambda=\SI{533}{nm}$.}
    \label{Fig:9}
  \end{center}
\end{figure}

Before further discussing the several interesting features that can be observed in the measurements reported in Fig.~\ref{Fig:8}, we present additional measurements obtained by the multimodal microscope in Fig.~\ref{Fig:9} for the same random array used in obtaining the results presented in Fig.~\ref{Fig:8}.  In particular, Fig.~\ref{Fig:9} shows results for polar angles of incidence $\theta_t=\ang{0}$, \ang{10}, \ang{20} and \ang{30} and without any angular restriction in the
collection arm of the microscope [\textit{i.e.} $\theta_t\leq\ang{45}$]. Again one finds that the intensity patterns measured with the microscope agree rather well with the corresponding  results obtained with the goniospectrophotometer.
It should be remarked that, located at the lower right edges of Figs.~\ref{Fig:9}(a)--(c), one can see a few irregular spots~(in yellow)  These features we believe to be due to  contamination present on the surfaces of the optical elements of the microscope and they are only visible when the intensity measurement requires an extremely high dynamic range. These irregular features are therefore artifacts and must not be considered as part of the light scattered by the sample.

The results presented in Figs.~\ref{Fig:8} and \ref{Fig:9} show explicitly that halos are present in the angular distribution of  the transmitted intensities for \textit{both} regular and random arrays of micropillars if $\theta_i\neq \ang{0}$. Furthermore, for both array types we find that the polar angle of transmission defining the halo is related to the polar angle of incidence by the relation $\theta_t=\theta_i$, which for our geometry, also is the polar angle of the specular direction of transmission. These results suggest that the presence of the halo is related to individual micropillars and their cylindrical shape rather than to how they are arranged on the surface of the substrate.

\smallskip
The origin of the halo can in fact be understood within the framework of either the extended Mie theory for non-spherical particles~\cite{Cohen1982} or the Debye series approach~\cite{Xu2010}. According to the later formalism, the origin of the halo can be attributed to rays which have been directly transmitted 
through the microsized cylinders (like in an optical fiber), or which have been directly reflected (or scattered) by the outer surface of the cylinders.
A detailed analysis of the halo and its origin will be presented in a separate publication using polarization dependent measurements. This study will be based on polarization sensitive measurements and it concludes that the halo is a consequence of reflection and/or transmission of light by \textit{individual} cylinders, and not due to, for instance, multiple reflections or the excitation of leaky guided modes in the glass slide (substrate).

\subsubsection{Concentric circular fringes}
\label{SubSec:3B3}

\begin{figure}[tbhp]
  \begin{center}
    \centering
    \includegraphics[width=0.85\columnwidth]{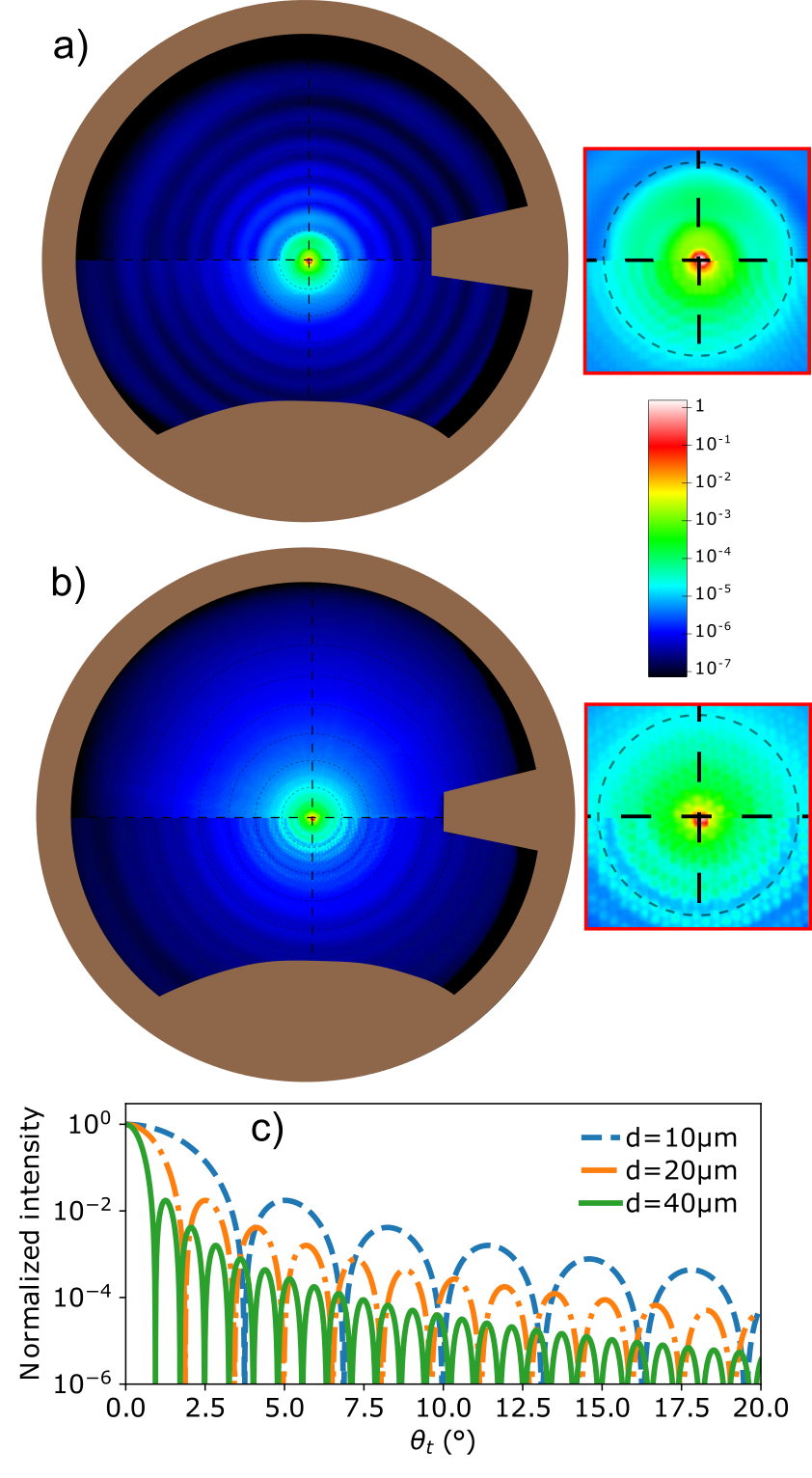}
    \caption{Angle-resolved DTCs for (a)~square or (b)~hexagonal arrays of supported micropillars obtained for normal incidence~[$\theta_i=\ang{0}$]. 
      The illumination was done with a Xenon lamp filtered at the center wavelength $\lambda=\SI{535}{nm}$ by a window of width \SI{10}{nm}.
      (a)~The diameters of the micropillars and the lattice constants of the square arrays were  $d=\SI{40}{\micro\meter}$ and $a=\SI{80}{\micro\meter}$ (upper half); and  
      $d=\SI{20}{\micro\meter}$ and $a=\SI{40}{\micro\meter}$~(lower half). (b) An hexagonal array of micropillars defined by  $d=\SI{10}{\micro\meter}$ and $a=\SI{30}{\micro\meter}$ measured with~(upper half) and without~(lower half) a black cover on both front and back sides. Each cover had a SI{3}{cm} hole in diameter that was big enough to not interact with the incident beam, and was aligned with the incident beam. (c) Airy patterns from Eq.~\eqref{eq:4} for circular apertures of diameter
      \SI{10}{\micro\meter}, \SI{20}{\micro\meter} and \SI{40}{\micro\meter}, respectively.}
   
    \label{Fig:10}
  \end{center}
\end{figure}

During the discussion of the results presented in  Figs.~\ref{Fig:5}(a)--(b), we observed two classes of concentric circular intensity patterns; \textit{class~1 fringes} are concentric about the direction defined by $\theta_t=\ang{0}$ while \textit{class~2 fringes} are centered around the direction of specular transmission $(\theta_t,\phi_t)=(\theta_i,\ang{0})$. The azimuthally symmetric circular fringes that are centered at $\theta_t=\ang{0}$ are readily observed in Fig.~\ref{Fig:10}(a) for sufficiently large values of $\theta_t$; these are the class~1 fringes for this configuration. In particular what is presented in Fig.~\ref{Fig:10}(a) are the angle-resolved DTCs for normal incidence obtained for two square arrays of micropillars where the micropillars of the first had diameter  $d=\SI{40}{\micro\meter}$~(upper half)  and the second $d=\SI{20}{\micro\meter}$~(lower half). The lattice constants were rather different for the two arrays and defined by $a=2d$. In both cases the heights of the micropillars  were $h=\SI{10}{\micro\meter}$. The surface coverage of square lattices is given by $\rho =(\pi/4)(d/a)^2$ and hence the two square arrays had the same surface coverage~[$\rho=\pi/16$]. A careful inspection and comparison of the results in the upper and lower halves of Fig.~\ref{Fig:10}(a) reveals that the (class~1) fringes for the two samples are rather similar even if the size of the  micropillars and the lattice constants were different for the two samples. Furthermore, when the illumination is oblique, the azimuthal symmetry of these fringes is lost, as seen in Fig.~\ref{Fig:8}, for instance, but the fringes remain centered at $\theta_t=\ang{0}$.It should be noted that class~1 fringes are also present in the intensity of the light transmitted through random arrays of micropillars~[Fig.~\ref{Fig:8}; upper halves]. In particular, for normal and oblique incidence, for random and regular arrays of micropillars, and for different sizes of micropillars, we find that the polar angle separation between two consecutive maxima of the fringes is of the order \ang{10}.

These results seem to rule out a specific size of the mircropillars, or the sample being periodic or random, as the main reason for the fringes pattern that we observe. Instead we believe that the origin of the class~1 fringes in the transmitted intensities can be attributed to guiding of light inside the sample. To support this interpretation we conducted an additional experiment, in which we remeasured the angle-resolved DTC for normal incidence for the sample used to produce the results in Fig.~\ref{Fig:5}~[hexagonal array of micropillars defined by $d=\SI{10}{\micro\meter}$ and $a=\SI{30}{\micro\meter}$]. However, in this additional experiment, the front and back sides of the sample were covered by a piece of black paper with the hole in the center each, where a hole of \SI{3}{cm} diameter was aligned with the incident beam on each side of sample. The diameter of the hole was chosen to be larger than the full width of the incident beam in order to allow the transmission of the direct beam, but, at the same time, partially block the light presumably partially guided inside the sample. By partially we mean that light is not perfectly confined inside the glass substrate as it would be in an ideal waveguide. The guided light modes in the glass are converted to radiative modes because the faces of the substrate are neither infinite nor perfectly smooth. Roughness and micropillars contribute to convert guided modes to radiative ones. The result for the angular distribution of the DTCs, with and without the black cover paper (on both the front and back side), are presented in Fig.~\ref{Fig:10}(b) as the upper half and lower halves, respectively. A comparison of these two data sets show clearly that the class~1 fringes vanish, or at least are significantly suppressed, when the sample has the black front and back covers. This finding we take as a direct confirmation that guiding of light in the sample is what causes the class~1 circular fringes.

\smallskip
Additional fringes, which are of another type than those we just discussed, are observed in the intensity transmitted through regular or random arrays of micropillars that are concentric about the direction of specular transmission. For instance, in the insets to Figs~\ref{Fig:8}(a) and \ref{Fig:10}(a), showing the details around the specular direction of transmission, they are seen for smaller values of $\theta_t-\theta_i$ (and $\phi_t-\phi_i-\ang{180}$), and these fringes are examples of the class~2 fringes. Note that such fringes are also seen in Fig.~\ref{Fig:5}(b). The class~2 fringes are in the measured data superimposed on the class~1 fringes so that it can be challenging to distinguish them in certain cases. However, the frequency of the class~2 fringes is typically found to be significantly higher than the frequency for the class~1 fringes. The class~2 fringes do depend on the size of the micropillars which contrasts what we found for the class~1 fringes. For instance, the upper and lower halves of Fig.~\ref{Fig:10}(a) compare DTCs for regular arrays of micropillars of different sizes (and lattice constants); the diameters of the micropillars were $d$=\SI{20}{\micro\meter} and $d$=\SI{40}{\micro\meter}, respectively. From the inset to this figure it is found that the polar angle separation between consecutive fringes close to the specular direction is smallest for the array consisting of micropillars of the largest diameter. Furthermore, the class~2 fringes, like their class~1 counterparts, can be observed for both regular and random arrays of micropillars. In Fig.~\ref{Fig:8}(b), the class~2 fringes are observed clearly as oscillations along the halo for both the random and regular array, and the angular distance between them are of the same order for both cases. These results we take as indications that the class~2 fringes are related to the size of the micropillars, not to how they are organized along the surface of the substrate.

The results for the class~2 fringes for normal incidence presented in Figs.~\ref{Fig:10}(a)--(b) we rationalize in the following way. The normally incident light of wavelength $\lambda$ is assumed to couple into the micropillars of diameter $d\sim\num{E1}\lambda$. This guided light will be radiated into the glass slide by the open-ended circular waveguide giving rise to an Airy like diffraction pattern predicted by Fraunhofer diffraction~\cite{BornWolf}. Taking a circular aperture, for simplicity, the
expression for the normalized transmitted intensity distribution of the diffracted light reads~\cite{BornWolf}
\begin{align}
  \bar{I}(\theta_t)
  &=
    \left(
    \frac{ 2 \, \mathtt{J}_1\!\!\left(\pi\frac{d}{\lambda}\sin\theta_t \right)
    }{
    \pi\frac{d}{\lambda}\sin\theta_t
    }
    \right)^2,
  \label{eq:4}
\end{align}
where $\mathtt{J}_1$ denotes the Bessel function of the first kind and order one, and $d$ denotes the diameter of the aperture that we will set equal the diameter of the micropillars~\cite{BornWolf}. Such normalized intensity distributions are shown on Fig.~\ref{Fig:10}(c) for diameter $d=\SI{10}{\micro\meter}$, \SI{20}{\micro\meter} and \SI{40}{\micro\meter}, respectively. The frequency of the oscillations observed in the transmitted intensities in the region around the specular direction of transmission are clearly different for the two samples considered in Fig.~\ref{Fig:10}(a), and the trend that one finds, is in agreement with the prediction of Eq.~\eqref{eq:4} [see also Fig.~\ref{Fig:10}(c)].  This demonstrates the high sensitivity of the angular position and intensity of the oscillations to the size of the cylinders, which is in good agreement with the predictions made on the basis of the Fraunhofer diffraction formalism. A direct comparison of the measurements and the Airy patterns is not straightforward for samples with a regular array of pillars, since the Airy pattern is modulated by the presence of diffraction orders.

\subsection{Haze and gloss}
\label{Sec:Haze}

Two of the most commonly used  integral optical measurements are haze and gloss. 
According to the two standards, ISO 14782:1999~\cite{ISO1} and ASTM D1003-13~\cite{ASTM1}, haze is defined as the percentage of
transmitted intensity, passing through a specimen, which deviates from the specular direction of transmission by  more
than 0.044 rad (\ang{2.5})~\cite{Simonsen2009}. Similarly, gloss is defined as the total intensity scattered inside a small angular region about the specular direction normalized by the intensity that is scattered by a standard sample; this ratio defines the ``gloss unit'' scale. According to the standards
ISO~2813:2014~\cite{ISO2} and ASTM~D523-14~\cite{ASTM2}, gloss can be measured at three different polar angles of incidence: $\theta_i=\ang{20}$, \ang{60} and \ang{80}~\cite{Simonsen2005}.

\begin{figure}[tbhp]
  \begin{center}
    \centering
    \includegraphics[width=0.99\columnwidth]{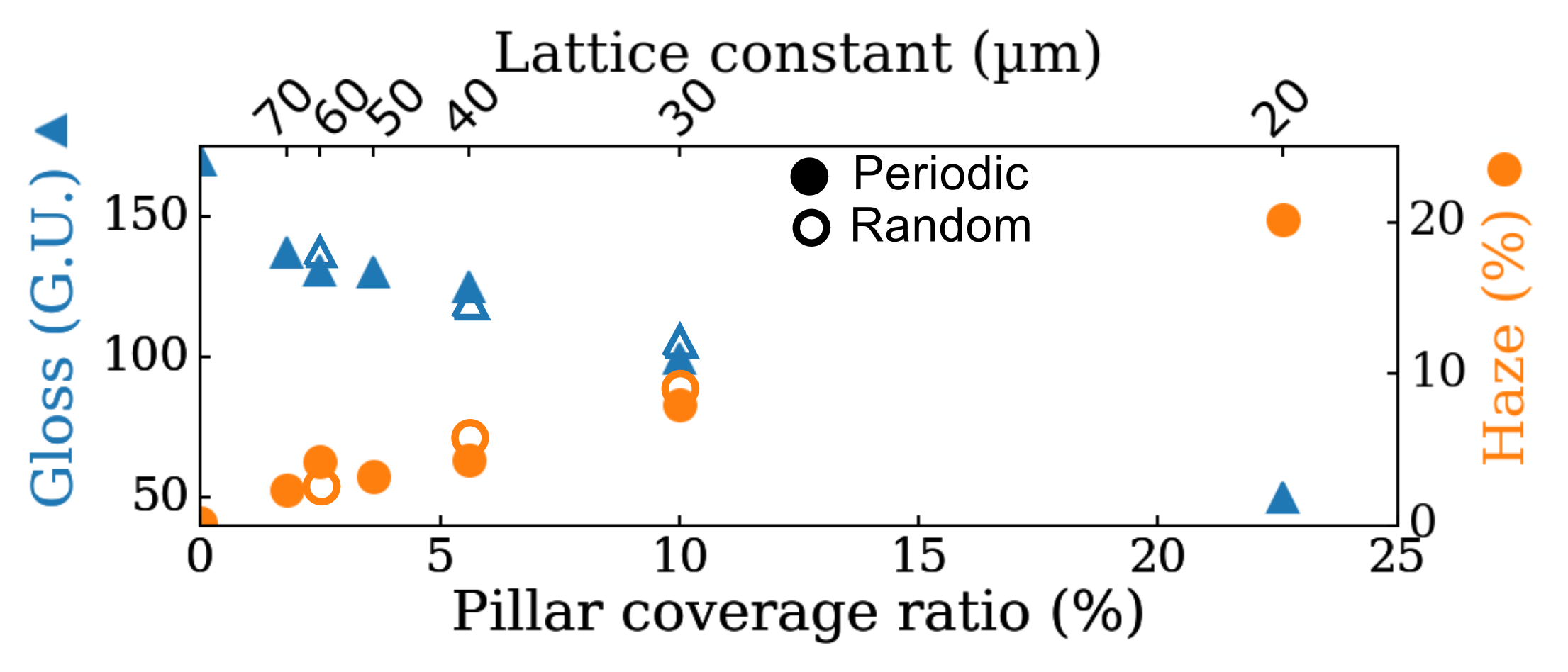}
    \caption{Measured haze and gloss values for hexagonal arrays of micropillars~(filled symbols) or random arrays of micropillars~(open symbols). The gloss values were measured with the glossmeter~(Enrichsen -- Pico Glossemaster Model 500) in reflection at $\theta_i=\ang{20}$. The haze values were obtained by a hazemeter~(BYK Gardner Haze-gard plus) in transmission at $\theta_i=\ang{0}$. The diameter and height of all micropillars were both \SI{10}{\micro\meter}.}
    \label{Fig:4}
  \end{center}
\end{figure}

We will now discuss the haze and gloss values that can be obtained for our supported regular and random arrays of micropillars. The gloss measurements that we report were performed in reflection for a polar angle of incidence of $\theta_i=\ang{20}$, while haze was measured in transmission for normal incidence~[$\theta_i=\ang{0}$].
Figure~\ref{Fig:4} presents the experimental values obtained for haze and gloss~(filled symbols) as functions of the surface coverage of the supported arrays by the use a hazemeter and a glossmeter. These results were obtained by keeping the diameter of the pillars constant at $d=\SI{10}{\micro\meter}$ and changing the lattice constant of the hexagonal array between $a=\SI{20}{\micro\meter}$ and \SI{80}{\micro\meter}.
We also performed measurements of haze and gloss for some samples of random arrays of micropillars [open symbols in Fig.~\ref{Fig:4}], and only minor differences  were found between values of these quantities obtained for regular and random arrays of the same surface coverage. From the results presented in Fig.~\ref{Fig:4} it is observed that the overall trend is that haze (in transmission) is an increasing (and approximately linear) function of the surface coverage while gloss decreases as 
function of the same quantity. The increase in haze with surface coverage indicates an increase in the amount of light transmitted away from the specular direction by more than \ang{2.5}. This observation is consistent with the decreasing level of measured gloss for the same samples. It is tempting to interpret this observation as a decrease in the specular component of the transmitted intensity with increasing surface coverage, while, at the same time, the diffuse component increases.

When the angular dependency  of the DTC is integrated over the full solid angle of transmission~[$\mathrm{d}\Omega_t=\sin\theta_t\,\mathrm{d}\theta_t\mathrm{d}\phi_t$], the transmittance of the sample is obtained~\cite{Hetland}. To calculate the haze of a sample from the angular distribution of the DTC measured for a given angle of incidence, one proceeds in the following way. First one performs a solid angle integration of the DTC over an angular region around the specular direction of transmission $(\theta_t,\phi_t)=(\theta_i,\phi_i-\ang{180})$ defined by $|\theta_t -\theta_i| \leq \Delta\theta_t$. For given angles of incidence, the haze of the sample is calculated by first dividing the result obtained in this way by the transmittance and then subtracting the resulting ratio from unity~(see Ref.~\cite{Simonsen2009}).
We have chosen  $\Delta\theta_t=\ang{2.5}$ to be consistent with the standard, and what is assumed in the construction of the hazemeter.

\begin{figure}[tbhp]
  \begin{center}
    \centering
    \includegraphics[width=0.95\columnwidth]{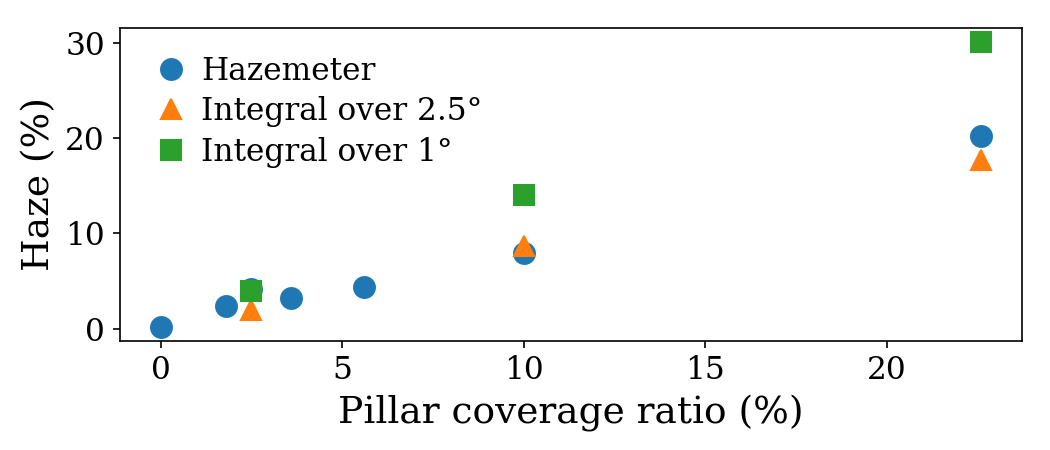}
    \caption{Comparison of the haze values for hexagonal arrays obtained with the hazemeter (direct measurements) and those calculated from angular resolved DTC data as functions of the surface coverage. The DTC data used in obtaining these results were measured at normal incidence by the use of the goniospectrophotometer. The measurement configuration was identical (accept for the angles of incidence) to what was used in obtaining the results in Fig.~\ref{Fig:5}(a). The way that haze was calculated from such data is described in the main text. The value of the polar angular interval  $\Delta\theta_t$  assumed in such calculations are given in the legend. It is noted that the value $\Delta\theta_t=\ang{2.5}$ corresponds to the norm of haze measurements. The diameter and height of all micropillars were both \SI{10}{\micro\meter}.}
    \label{Fig:7}
  \end{center}
\end{figure}

Figure~\ref{Fig:7} compares the haze values calculated in this way from the angle-resolved DTC measurements with the values obtained by direct haze measurements performed on the same samples of hexagonal arrays in transmission and at normal incidence; in such calculations, the missing data points were set to zero. One finds good agreement between haze values measured directly and values calculated from the measured angle-resolved DTC data. This testifies to the consistency of the angle-resolved measurements and their normalization. For instance, the haze values calculated from the angle-resolved DTC data are \SI{18}{\percent} for lattice constant a=\SI{20}{\micro\meter} (or surface coverage $\rho$=\SI{23}{\percent}); \SI{9}{\percent} for a=\SI{30}{\micro\meter} [$\rho$=\SI{10}{\percent}]; and \SI{2}{\percent} for a=\SI{60}{\micro\meter} [$\rho$=\SI{3}{\percent}]. These values should be compared to the corresponding measured haze values for the same samples which are \SI{20}{\percent}, \SI{8}{\percent} and \SI{4}{\percent}, respectively. An increase of the lattice constant $a$, or equivalently, a decrease of the surface coverage $\rho$, cause more of the diffractive orders to end up inside a cone of angular width $\Delta\theta_t$ about the specular direction, and hence, the value of haze to drop. Yet, transmission efficiencies of individual diffraction orders are naively expected to decay with increasing lattice constant and increasing order. Therefore, for sufficiently large lattice constant, a further increase of it will only marginally affect the resulting haze value. However, for smaller lattice constants, for which only a few diffractive orders fall inside $\Delta \theta_t < \ang{2.5}$, this is no longer the case. Therefore, from haze values measured with a hazemeter alone, we are not able to distinguish a regular array from a random array if the surface coverage is sufficiently large. For reasons of comparison, Fig.~\ref{Fig:7} also reports results for the calculations assuming $\Delta\theta_t=\ang{1}$ a value suggested as more realistic in a recent study on the angular width of specular beams~\cite{Leloup2016}. As expected, when using this value for $\Delta\theta_t$, the calculation procedure results in values that are higher than the haze values for the same sample. However, the interesting observation is not that one gets larger values, but how much larger the obtained values are. For small values of the surface coverage, or larger lattice constants, the difference between the results obtained when using these two values for  $\Delta\theta_t$ in the calculation of haze are not very dramatic. However, for the larger values of the surface coverage the differences increase. 

\smallskip
When the visual aspect of an object is to be taken into account, the particularities of the human eye, rather than those of a artificial detector, must be considered. For instance the angular resolution of the human eye is about \ang{0.03}~\cite{Zettler1976}.
If haze is intended to quantify the fraction of transmitted intensity that is transmitted away from the specular direction (of transmission), the use of the value $\Delta \theta_t=\ang{2.5}$, which is perfect for a number of industrial needs is far too large when it comes to discuss human visual perception of the optical response of 1D or 2D gratings.

\begin{figure}[tbhp]
  \begin{center}
    \centering
    \includegraphics[width=0.9\columnwidth]{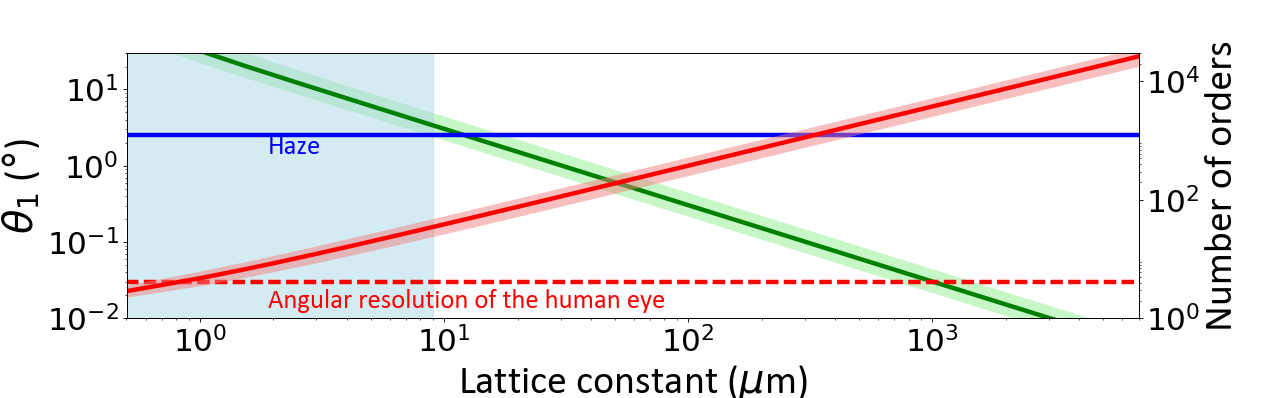}
    \caption{Normally incident light~[$\theta_i=\ang{0}$] diffracted through a surface consisting of a one-dimensional grating of lattice constant $a$. The green line shows the angle of diffraction $\theta_1$ of the first diffraction order [$m=1$], while the red line represents the total number of propagating diffracted orders $N$, both obtained by assuming the wavelength $\lambda=\SI{500}{nm}$ for the incident light. The corresponding green and red shaded areas (around the solid lines of the same color) represent  the variations of these two quantities due to the wavelength of the incident light varying over the visible range \num{380}--\SI{780}{nm}. The blue horizontal line corresponds to the smallest diffraction angle~[\ang{2.5}] for which the diffracted light in transmission contributes to haze. Haze measurements for lattice constants smaller than approximately \SI{9}{\micro\meter} receive contribution from all higher order diffractive orders for which $m\neq 0$. However, for larger lattice constants not \textrm{all} such higher orders will contribute. 
      For comparison, the horizontal red dashed line corresponds the limit of angular resolution of the human eye.
    } 
    \label{Fig:1}
  \end{center}
\end{figure}

 For the sake of illustration, let us restrict ourselves to a one-dimensional grating at normal incidence for which the polar angle of the propagating diffractive orders in transmission is given by $\sin\theta_m=m(\lambda/a)$ with $m$ and an integer ($m\in\mathbb{Z}$) and $m=0$ corresponds to specular transmission. For this system, the number of propagating diffracted orders in transmission is (red curve in Fig.~\ref{Fig:1}) $N = 2 \left \lfloor{\frac{a}{\lambda}}\right \rfloor  + 1$,
where $\left \lfloor{x}\right \rfloor$ denotes the floor function of $x$ which returns the greatest integer less than or equal to its argument $x$. At normal incidence the two first diffractive orders are symmetric about the specular direction and correspond to the polar angles of transmissions $\pm\theta_1$.
Figure~\ref{Fig:1} illustrates the variation of the angle of the first diffracted order $\theta_1$ under the assumption of normal incidence [$\theta_i=\ang{0}$]. The green solid line in this figure corresponds to an illumination  wavelength of $\lambda=\SI{500}{nm}$ while the green area around this line  represents the variation due to the whole visible range of wavelengths from \SIrange{380}{780}{nm}. Moreover, the red solid curve in Fig.~\ref{Fig:1} illustrates the total number of diffraction orders in such system.

For a lattice constant of about a=\SI{9}{\micro\meter} and an illuminating  wavelength  of $\lambda=\SI{380}{nm}$, the first diffracted order will enter into the ``specular'' area of the haze measurement. Starting from this wavelength (and higher) the haze measurements, as a measure of the fraction of transmitted intensity away from the specular direction,  are biased (unshaded region of Fig.~\ref{Fig:1}). For lattice constants all the way up to $a=\SI{1}{mm}$ a human observer will be able to distinguish specular transmission from the first diffraction order. These results hints towards a not optimal definition of haze for gratings of long  periods compared to the wavelength of visible light. According to our discussion, when the visual aspect of gratings with large periods matters, a haze definition making use of smaller angular spread around the specular direction than the actual 2.5° defined in the norms, would provide objective haze values which would be in agreement with the subjective experience of the human eye.

\section{Conclusions}
\label{Sec:Concusions}

We report experimental results for the angle-resolved transmitted intensity measurements for a set of regular or random arrays of dielectric micropillars in the low coverage limit that are supported by thin index matched glass slides. The regular arrays were characterized lattice constants in the range from $a=\SI{20}{\micro\meter}$ to \SI{80}{\micro\meter}. The measurements were performed by either a goniospectrophotometer or a multimodal imaging polarimetric microscope and the two sets of measurements gave comparable results. On the basis of experimental data obtained in this way, it is demonstrated that for identical micropillars, the mean differential transmission coefficients for the random arrays agree well with the envelope of the same quantity for the regular array under the assumption that the surface coverage is the same. Moreover, we find that the angle-resolved measurements display unique diffractive features that are due to properties of single micropillars and not to how they are organized along the surface. Finally we perform a comparison of direct measurements of haze in transmission for our structured samples with what can can calculated from the angle-resolved transmitted intensity measurements. Good agreement between the two types of results are found which testifies to the accuracy of the angle-resolved measurements that we report. However, we find that for larger surface coverage, haze values alone can not be used to distinguish regular and random arrays of micropillars.

\bigskip
\textbf{Acknowledgments}
We thank Gael Obein, Guillaume Ged, Sebastien Noygues and Emmanuel Garre for valuable discussions.

French National Research Agency (ANR-15-CHIN-0003; IDEX Paris-Saclay ANR-11-IDEX-0003-02)

\end{document}